\documentclass[aps,pre,twocolumn,showpacs]{revtex4}
\usepackage{graphicx}%
\begin{document}
\title{Thermodynamics of force-induced B-DNA melting: single-strand discreteness matters}

\author{Nikos Theodorakopoulos$^{1,2}$ } 
\affiliation{
$^{1}$Theoretical and Physical Chemistry Institute, National Hellenic Research Foundation,
Vasileos Constantinou 48, 116 35 Athens, Greece \\
$^{2}$Fachbereich Physik, Universit\"at Konstanz, 78457 Konstanz, Germany\\
}
\date{\today}
\begin{abstract}
\pacs{36.20.-r,  87.14.Gg, 87.15.Aa}
Overstretching of B-DNA is currently understood as force-induced melting. Depending on the geometry of the stretching experiment, the force threshold for the overstretching transition is around  65 or 110 pN. Although the mechanisms behind force-induced melting have been correctly described by Rouzina and Bloomfield \cite{RouzinaBloomfield2001a}, neither force threshold has been exactly calculated by theory. In this work, a detailed analysis of the force-extension curve is presented, based on a description of single-stranded DNA in terms of the discrete Kratky-Porod model, consistent with (i) the contour length expected from the crystallographically determined monomer distance, and (ii) a high value of the elastic stretch modulus arising from covalent bonding. The value estimated for the ss-DNA persistence length, $\lambda = 1.0 $ nm,  is at the low end of currently known estimates and reflects the intrinsic stiffness of the partially, or fully stretched state, where electrostatic repulsion effects are expected to be minimal. A detailed analysis of single- and double-stranded DNA free energies provides estimates of the overstretching force thresholds. In the unconstrained geometry, the predicted threshold is   64 pN. In the constrained geometry, after allowing for the entropic penalty of the plectonemic topology of the molten state, the predicted threshold is 111 pN.

\end{abstract}
\maketitle


\section{Introduction}
Understanding the elastic properties of single-stranded (ss) DNA is an important problem in a variety of biophysical contexts, e.g.  molecular recognition,  folding into hairpin-like structures, unzipping and/or overstretching of the double helix. The latter process, where an applied force of about 65 pN (or 110 pN, depending on the boundary conditions) destroys the B-DNA structure   \cite{Busta1996} is - following some speculation over a possible distinct double-stranded (ds) intermediate state - currently understood as force-induced melting \cite{vanMam2009, RouzinaBloomfield2001a}. 

Current estimates of ss-DNA flexibility, as expressed by the persistence length, vary, mainly according to the salinity of the solution, between 0.75 and 2 nm \cite{Chen2012, LipfertDoniach2012, Ritort2014, Roth2018}. The  lowest value corresponds to a theoretical limit of high salt concentration which acts to shield out any electrostatic repulsion and therefore reflects in some sense the intrinsic stiffness of the bonds which form the polymer chain. On the other hand, the distance between successive monomers has been determined by a careful averaging of available crystallographic data to be  $0.63 $ nm \cite{Murphy2004}, i.e. comparable to the persistence length. This renders the use of a continuum approximation, and hence the use of the wormlike chain (WLC) model in analyzing ss-DNA flexibility data rather questionable - although in fact many of the present estimates have been based upon this model.

In this paper, I analyze the force-extension data on DNA overstretching \cite{vanMam2009} in terms of the discrete Kratky-Porod (KP) model \cite{Kratky-Porod}. The analysis allows for elastic stretching of ss-DNA beyond its contour length, using an estimate of the stretch modulus based on {\em ab initio} calculations \cite{GaubNetz2005}, which is in line with AFM measurements \cite{Gaub2000, GaubNetz2005}. 

The data can be well described in terms of a monomer distance equal to 0.63 nm, the average crystallographic value, and a persistence length equal to $1.0 $ nm. This value of the persistence length lies at the low end of current estimates and is consistent with the physics of the stretched state, which is expected to be less responsive to the electrostatic repulsion, and therefore reflects the intrinsic entropic stiffness of the chain.

I present a simple thermodynamic analysis of the overstretching transition, along the lines developed in \cite{RouzinaBloomfield2001a}. In the case of unconstrained geometry,  using the elastic energies of ss- and ds-DNA, along with standard DNA melting parameters as the only input, I calculate the threshold of the overstretching force, which turns out to be  in excellent agreement with the experimentally determined one. The case of constrained geometry is slightly more complicated. The ends of both strands of the DNA molecule are held fixed. This means that the molten state does not release the full amount of entropy corresponding to two free chains. Assuming that, in the overstretched state,  the two ss-DNA strands wrap around each other with a linking number \lq\lq{}inherited\rq\rq{} from the original, double-helical structure  \cite{vanMam2009}, one can calculate the entropic penalty arising from the constraint.  Taking this into account predicts a force threshold very close to the observed one and supports the view of the plectonemic  structure of the overstretched state.

The paper is divided in four parts. Section II provides some necessary background material on force-extension relationship, based on both  the KP model and its continuum WLC limit, and a comparative analysis of the two approaches in the case where the continuum approximation breaks down. Section III compares the results of theoretical calculations with experiment. A brief discussion of the results is presented in Section IV.

\section{The force-extension relationship}
\subsection{Kratky-Porod, model essentials}
The KP model 
describes chain polymers with nonzero rigidity.  A conformation of the chain with segments $j=1,\cdots N$, each of which has a fixed length $a$, and a direction vector ${\vec t}_j$ has a total elastic energy
\begin{equation}
\label{eq:DiKP}
	H = -\frac{\kappa}{a} \sum_{j=1}^{N-1} ({\vec t}_{j}\cdot{\vec t}_{j+1} -1)  - a\vec{f}\cdot  \sum_{j=1}^{N} {\vec t}_{j}\quad,
%
\end{equation}
where $\kappa $ is a measure of the chain\rq{}s stiffness and ${\vec f}$ represents an externally applied force. 
The stretched chain has a fixed contour length $L_0 = Na$. The WLC elastic energy can be obtained from (\ref{eq:DiKP}) in the limit
 $ a  =L_0/N, N \to \infty$, as
\begin{equation}
\label{eq:WLC}
{\hat H} = \frac{\kappa}{2}\int_0^{L_0} \> ds \left|\frac{\partial \vec t}{\partial s}  \right|^2
- {\vec f}\cdot \int_0^{L_0} \> ds \> {\vec t}(s)   \quad,
\end{equation}
 in terms of the continuous direction field ${\vec t}(s)$ and the local curvature.
 
 The canonical partition function in the presence of the external force is
 \begin{equation}
\label{eq:ZKPf}
Z_N(f) = \int d\Omega_1 \cdots d\Omega_{N} \prod_{j=1}^{N-1} e^{b({\vec t}_{j}\cdot{\vec t}_{j+1} -1)}  \> \prod_{j=1}^{N} e^{\beta a 
{\vec f} \cdot {\vec t}_{j} } \quad,
\end{equation}
where $\beta = 1/(k_B T)$, $k_B$ is the Boltzmann constant, $T$ the temperature, $b=\beta\kappa/a$ and $\Omega_j \equiv (\theta_j, \phi_j)$ the solid angle specifying the orientation of ${\vec t}_j$. 

The persistence length, defined as the characteristic scale at which orientational correlations decay, is given by
\begin{equation}
\lambda
	= -\frac{a}{ \ln \left(  {\rm coth} b - \frac{1}{b} \right) } \quad.
\label{eq:KPpersl}
\end{equation}

The quantity of interest is the average extension
 \[ 
 L  = <(\vec{R}_N -\vec{R}_0)\cdot {\hat f}> = \frac{1}{\beta}\frac{\partial}{\partial  f}
 \ln {\hat Z}_N (f)  \quad,
\]
where
\begin{equation}
\label{eq:Zfr}
{\hat Z_N}(f)= \frac{Z_N (f)}{Z_N(0)}  \quad,
\end{equation}
can be computed by using in (\ref{eq:ZKPf})  the standard expansion of the isotropic interaction in terms of spherical harmonics
\begin{equation}
 e^{b\vec{t}_j\cdot\vec{t}_{j+1} }= \sum_{l=0}^{\infty} \sum_{m=-l}^{l} i_l(b) \>Y_{lm}(\Omega_j) Y_{lm}^{*}(\Omega_{j+1}) \quad,
 \label{eq:ExpSph}
\end{equation}
where $i_l$ is the modified spherical Bessel function of $l$th order. The result of performing the angular integrations in  (\ref{eq:ZKPf}) can be expressed as the leading diagonal element of the matrix product
\[
{\hat Z_N}(f) =  ({\bf U}^N)_{00} 
\]
with the 
elements of the real, symmetric matrix ${\bf U}$ given by \cite{Errami2007, Marko2005}
 \begin{equation}
U_{ll^\prime} \equiv \frac{1}{2}[\hat i_l(b)\hat{i}_{l^\prime}(b)(2l+1)(2l^\prime+1)]^{1/2} F_{ll^\prime}
  \>, l,l^{\prime}=0,1,\cdots \>,
\label{eq:tll}
\end{equation}
where $\hat i_l(b)=i_l(b)/i_0(b)$ are ratios of the  modified spherical Bessel functions,
the $P_l$\rq{}s are Legendre polynomials with the normalization $P_l(0)=1$,
\begin{eqnarray}
	F_{ll'}({\tilde f})   	&	= & \int_{-1}^{1} \> dx P_l(x)P_{l^\prime}(x)e^{ {\tilde f} x}  \\ \nonumber
	 &= &\sum_{k=|l-l'|,k+l+l'=2r}^{l+l'}  (2k+1) (-1)^k
	    \frac{1}{r+1/2}\\  \nonumber
	  &   & \cdot  \frac{\Psi(r-k)\Psi(r-l)\Psi(r-l')}{\Psi(r)} 
	    i_k({\tilde f})  \quad,
\label{eq:fllclosed}	    
\end{eqnarray}
\begin{equation}
	\Psi(n) = \frac{\Gamma(n+\frac{1}{2} ) }  {\Gamma(n+1)\Gamma(\frac{1}{2})} =
	\prod_{j=1}^{n}\left( 1 - \frac {1}{2j}\right) \quad,
\end{equation}	
and ${\tilde f }=\beta a f$.
 
  The above equations provide a basis for a full numerical calculation of the force-extension curve in the case of discrete KP chains. If the number of monomers $N$ is large,  (\ref{eq:Zfr}) will be dominated by the largest eigenvalue, $\mu_0$, of the matrix ${\bf U}$. The force-dependent part of free energy will be given by 
 \begin{equation}
 G =  -N  k_B T \ln \mu_0    \quad,
 \label{eq:FreeEnKP}
 \end{equation}
  and the average extension, as a fraction of the contour length, will be
 \begin{equation}
 \frac{ L}{L_0} =  \frac{ \partial \ln \mu_0 }{\partial {\tilde f} }  \quad.
 \label{eq:FEKP}
 \end{equation}
 
  Numerical diagonalization of the infinite matrix  ${\bf U}$ necessitates its truncation to finite dimension $l_{max}$. In the cases considered in the present paper an $l_{max}$ of the order of 10  is quite sufficient, leading to very  rapid computation. As a practical matter, it is convenient to divide all elements $F_{ll^{\prime}}$ by an overall factor $i_0({\tilde f })$ and add an extra term 
 \[
 \ell_0( {\tilde f })= \coth {\tilde f } -1/{\tilde f }
 \]
 to the right hand side of (\ref{eq:FEKP}).
 
 \subsection{The WLC limit}
 In order to obtain the force-extension curve in the continuum (WLC) limit, we keep terms of order $a = L_0/N$ and let $N \to \infty$. There are two contributions of this order. The first comes from the asymptotic form of the  modified spherical Bessel functions \cite{Abramowitz} for large arguments $b=\beta \kappa /a$,
 \[
 i_l(b) \sim \frac{1}{2b} e^b \left\{ 1-\frac{l(l+1)}{2b} +{\cal O}(1/b^2)  \right\}
 \]
 resulting in
 \[
 {\hat i}_l(b) \sim  1-\frac{l(l+1)}{2b} 
 \]
 and generates a contribution of order $a$ from the prefactor in (\ref{eq:fllclosed}). The second  comes from expanding the exponential $e^{\beta afx}\approx 1 + a \beta f x$. Combining the two results in 
  \[
  {\bf U} ={\bf I} - \frac{1}{N} \frac{L_0}{\lambda}{\bf J} \quad,
\label{eq:TJ}
 \]
 where ${\bf I}$ is the unit matrix,
\begin{equation}
\label{eq:Jmelem}
J_{ll^\prime} =  
\frac{l(l+1)}{2} \delta_{l l^\prime} - \beta \lambda f  
 \frac{ l^\prime \delta_{l^\prime ,l+1}+ l \delta_{l^\prime ,l-1}}{[(2 l+1)(2 l^\prime+1)]^{1/2}}
    \quad,
\end{equation}
and, $\lambda =\beta \kappa$, the WLC persistence length. It is now possible to take the limit
 \[
 \lim_{N \to \infty}  {\bf U}^N = \lim_{N \to \infty} \left( 1-\frac{1}{N} \frac{L_0}{\lambda}{\bf J} \right)^N = e^{-L_0/\lambda {\bf J}}   \quad.
\]
Using the eigenvector expansion of the {\bf J} matrix results in 
   \begin{equation}
   {\hat Z}_{WLC} (f)=\lim_{N \to \infty} {\hat Z}_N(f) =  \sum_{\nu} e^{-L_0/ \lambda \Lambda_\nu}  \left| A_{\nu}^{0}\right|^2   \quad ,
   \label{eq:Zeigsum}
   \end{equation}
 where $\Lambda_\nu$  is the $\nu$th eigenvalue and $A_{\nu}^{0}=  <\nu|0>$ is the $l=0$ component of the $\nu $th eigenvector. The extension can, in general, be obtained by differentiating with respect to $f$. If the chain is flexible, i.e. the contour length $L_0$ is much larger than the persistence length $\lambda$,  (\ref{eq:Zeigsum})  will be dominated by the smallest eigenvalue $\Lambda_0$. In this case, the force-dependent part of the free energy will be
 \begin{equation}
 G  \approx k_B T  \frac{L_0 }{\lambda } \Lambda_0 \quad  {\rm if} \>  L_0\gg \lambda \quad,
 \label{eq:FreeEnWLC}
 \end{equation}
 and the relative extension
  \begin{equation}
 \frac{ L}{L_0} \approx - \frac{ \partial  \Lambda_0 }{\partial {\bar f} }  \quad {\rm if} \> L_0\gg \lambda \quad,
 \label{eq:FEWLC}
 \end{equation}
 where ${\bar f} = \lambda \beta f$. Note that the above  condition is always satisfied in the case of long, genomic samples, e.g. such as used in the overstretching experiments considered here. On the other hand, it is clearly violated by  DNA chains at the 100 nm scale, where  contributions from the whole eigenvalue spectrum must be taken into account.

 \subsection{The freely-jointed chain (FJC) limit}
 For the sake of completeness, I also consider the limit of a freely-jointed chain with a segment length (Kuhn length) equal to $2\lambda$. The relative extension is then given by
 \begin{equation}
 \frac{ L}{L_0} = \ell_0 (2 \beta \lambda f ) .
 \label{eq:FEFJC}
 \end{equation}
 
 \subsection{Enthalpic corrections}
As the applied force increases beyond a few pN, standard mechanical restoring forces come into play. In the case in double-stranded DNA, $\pi$-electron stacking forces provide a common physical origin of the stretching and bending modulus; enthalpic and entropic  elasticity are both important near the fully stretched state.  The estimated \cite{BustaCurrOp2000,Marko2003} stretch modulus is about $F_0=1000$ pN. 
 
For covalently bonded polymers this threshold may not set in until very high forces are applied. Early AFM measurements performed in the case of ss-DNA, which is bound by covalent  bonds, show that it can sustain forces at least up to 800 pN, with an elastic stretch of the order of 15\% beyond the crystallographic monomer distance \cite{Gaub2000}. This suggests an elastic stretch modulus of the order of 5 nN, implying a correction of the order of 1\% in the context of a 100 pN experiment. Further work combining AFM measurements with {\em ab initio} calculations \cite{GaubNetz2005} estimates a value of the elastic modulus $F_0= 8.4$ nN for ss-DNA.  I will use this estimate in the present analysis.
  
 Elastic stretching can be explicitly taken into account in the form
   \begin{equation}
 \left(\frac{ L}{L_0}\right)_{total} = \frac{ L}{L_0} \cdot \left(  1+ \frac{f}{F_0} \right)  \quad,
 \label{eq:stretch}
 \end{equation}
 where the first factor may originate from any of the entropic elasticity alternatives (KP, WLC, FJC).
 
\subsection{Why discreteness matters}
 I will now compare the force-extension curves provided by three different models describing a flexible system with a large number of monomers (N=1000), and monomer distance $a$ comparable to the persistence length. This corresponds closely to the situation of stretching experiments with ss-DNA.

\begin{figure}[h!]
\includegraphics[width=0.4\textwidth]{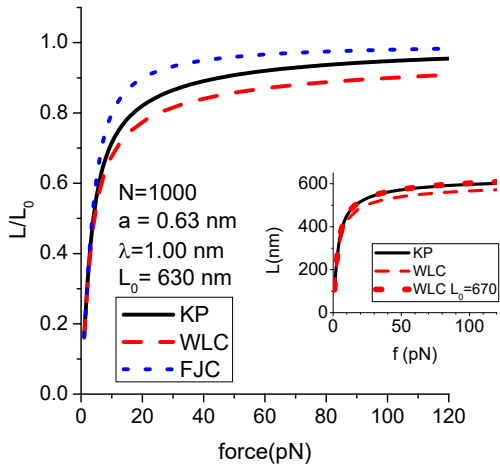}
\vskip -1.5truecm
\caption{The force-extension curve calculated for three different models in the case where the persistence length $\lambda=1.0 $nm, is comparable i to the monomer distance $a=0.63$ nm. The solid line is the KP calculation, according to (\ref{eq:FEKP}), the dashed line a WLC calculation according to  (\ref{eq:FEWLC}) and the dotted line the result of the FJC. The inset displays results for the same KP and WLC calculations on an absolute extension scale, where the contour length is 630 nm for both; the third curve, a dotted line, is the result of a WLC calculation with the same persistence length and an increased contour length, 670 nm, corresponding to a monomer distance $a=0.67$ nm, which successfully mimicks the force-extension curve of the shorter KP chain.
}
\label{fig:modelcomp}
\end{figure}
 
The first model is the Kratky-Porod model  with $N=1000$ segments, a monomer distance $a=0.63 $ nm and a stiffness constant $\kappa /a = 8.0 $ pN nm, which from (\ref{eq:KPpersl}) implies $\lambda=1.0 $ nm at a temperature 20 C. The second model is a WLC  with the same persistence length $\lambda = 1.0 $ nm and a contour length $L_0 = 1000 \times 0.63 $ nm $=630 $ nm. The third model is a freely-jointed chain (FJC) with a Kuhn length equal to $2\lambda=2.0 $ nm and a number of segments $L_0/(2\lambda)$. 

Fig. \ref{fig:modelcomp} displays the force-extension relationship in all three cases. The difference between the KP and WLC cases is fairly significant, of the order of 7\% at 70 pN. More importantly however, from a data analysis point of view, use of the  WLC enforces a bias towards higher contour length, i.e. monomer distances systematically larger than those mandated by crystallographic data (cf. inset).




\section{Overstretching of DNA}
 
\subsection{The force-extension curve}
\label{sec:DNAfe}
\begin{figure}[h!]
\includegraphics[width=0.4\textwidth]{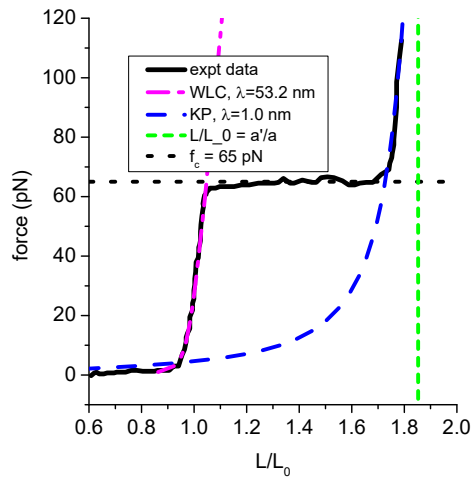}
\vskip -1.5truecm
\caption{The DNA force-extension curve at forces up to 120 pN. Extension is measured relative to the ds-DNA contour length. Continuous line, experimental data redrawn from  \cite{vanMam2009}. Dashed-dotted curve, theoretical WLC calculation with $\lambda=53.2$ nm, contour length $8400 \times 0.34 $ nm and elastic (stretch) modulus $F_0 =1000$ pN. Dashed curve, discrete Kratky-Porod calculation with $N=8400$, monomer distance $a^{\prime}=0.63 $ nm, stiffness constant  $\kappa / a^{\prime} = 8 $ pN nm, corresponding to $\lambda = 1.0 $ nm, and elastic stretch modulus 8.4 nN \cite{ GaubNetz2005}. A dotted horizontal line at 65 pN denotes the overstretch threshold. A vertical, short-dashed line shows the limit of ss-DNA contour length.}
\label{fig:overstretch}
\end{figure}

Fig.  \ref{fig:overstretch} displays the DNA force-extension curve as measured in an unconstrained geometry, i.e. where the chain is held fixed at the 3\rq{}-3\rq{} opposite ends \cite{vanMam2009}. The extension is defined relative to the contour length of ds-DNA.  

At low force levels, the data can be well described by the WLC model (\ref{eq:FEWLC}), corrected for elastic stretching according to (\ref{eq:stretch}) with an elastic (stretch) modulus $F_0 =1000$. The WLC calculation uses a contour length $L_0 =  Na$ computed with $a= 0.34 $ nm and $N=8400$ \cite{vanMam2009} and a persistence length $\lambda = 53.2 $  nm. Only the latter parameter has been adjusted in order to obtain a visual fit to the data; its value is in excellent agreement with the one obtained from earlier force-extension measurements \cite{Busta1994}.

At forces beyond the 65 pN plateau, the data can be fitted to the KP model force-extension relationship (\ref{eq:FEKP}) for ss-DNA, corrected for elastic stretching according to (\ref{eq:stretch}). I have used  the monomer distance $a^{\prime}=0.63 $ nm, as determined by averaging over crystallographic data \cite{Murphy2004}, the number of bases $N=8400$ as given in \cite{vanMam2009}, and a stretch modulus equal to 8.4 nN \cite{GaubNetz2005}. Again, the only adjustable parameter used to obtain a visual fit is the local stiffness $\kappa/a^{\prime}= 8.0 $ pN nm, which corresponds to a persistence length  $\lambda = 1.0 $ nm. 

\subsection{Thermodynamics (unconstrained geometry)}
The plateau in the force-extension curve characterizes the coexistence of the two phases, double-stranded and single-stranded. In this geometry only one of the two single strands of the molten phase is under an applied force.  Coexistence occurs when the  difference in elastic free energies between single- and double-stranded phases fully compensate the stability  free energy of the force-free double helix. 

\subsubsection{Elastic free energies}

The force-dependent parts of the elastic free energies per unit are 
\begin{equation}
\label{eq:g_ss}
g_{ss}(f) = -k_B T \ln \mu_0  +  \frac{1}{2} \frac{a^{\prime}}{F_0} f^2 
\end{equation}
and 
\begin{equation}
\label{eq:g_ds}
g_{ds}(f) = k_B T \frac{a}{\lambda} \Lambda_0   +  \frac{1}{2} \frac{a}{F_0} f^2 
\end{equation}
for the single- and double-stranded cases, respectively, where the second term expresses the elastic stretch energy.

\subsubsection{Thermal stability of the double helix}

The  double helix is generally stable at room temperature. This is usually expressed in terms of the free energy difference per base pair between duplex and molten state
\begin{equation}
\label{eq:dhstab}
\Delta G = - (\Delta H - T \Delta S) = ( T-T_m)  \Delta S \quad, 
\end{equation}
where $\Delta H>0$ and  $\Delta S>0$  denote, respectively,  average values for the energy and the entropy of dissociation per base pair and  $T_m=\Delta H/\Delta S$ is the melting temperature. A detailed analysis of DNA melting thermodynamics \cite{Blake91} suggests that the constant value  $\Delta S \approx 12.5     \>  k_B$ consistently describes, in a mean-field fashion, the cooperative melting behavior of a large number of samples of varying sequence and composition.The above value has been used in \cite{RouzinaBloomfield2001a}  and will also be used here.

The melting temperature $T_m$ can be calculated from the empirical Marmur-Schildkraut-Doty \cite{MarmurDoty1960,SchildkrautDoty1962,DelcourtBlake1998} equation 
\begin{equation}
T_m (C) = 193.67 - (3.09-x)\cdot (34.47- 6.52 \log_{10}[c])  \quad,
\end{equation}
where $x$ is the  GC fraction and $c$ the salt concentration. At $x=0.5$ and $c=150$  mM, this leads to $T_m=90.4 $ C. It should be noted that (\ref{eq:dhstab}) accounts for all effects involved in the relative stability of ds- and ss-DNA, including entropic polymer contributions, except those explicitly originating in the applied force and described by (\ref{eq:g_ss}) and  (\ref{eq:g_ds}). 

\subsubsection{Overstretching threshold}
\begin{figure}
\includegraphics[width=0.4\textwidth]{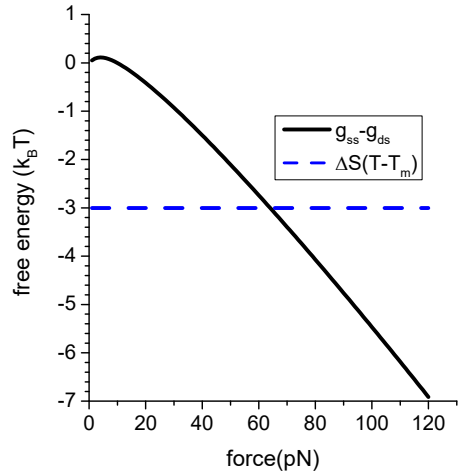}
\vskip -1.5truecm
\caption{Free energy balance resulting in DNA overstretching (unconstrained case).The solid line denotes the difference in elastic free energy between single- and double-stranded state. The dashed line denotes the duplex stability free energy at 20 C. Free energies are in units of $k_B T$. The intersection defines the overstretch threshold at 64 pN.
}
\label{fig:overstretch1}
\end{figure}

Fig. \ref{fig:overstretch1} displays the free energy difference $g_{ss}(f) - g_{ds}(f)$ at $T=20$ C. After an initial positive phase at very low forces ($ < 10$ pN), the difference becomes increasingly negative, until, at 64  pN it compensates fully for the room temperature duplex stability (\ref{eq:dhstab}). This marks the onset of the overstretching transition.

\subsection{Thermodynamics (constrained geometry)}
In the constrained geometry version of the overstretching experiment all four ends of the 3\rq{}5\rq{}-5\rq{}3\rq{} chain are attached; each single strand is  subjected to half the external force. The elastic free energy difference is now $2 g_{ss}(f/2)-g_{ds}(f)$. According to the discussion of the previous subsection, overstretching should occur at the force level where the elastic free energy difference fully compensates duplex stability. 

The geometry of the constraint however prohibits a full entropy release in the overstretched state. The two strands remain stretched, attached at both ends. Moreover, available conformational  space is further restricted by the entanglement brought about by the previous double helical order. An \lq\lq{}inherited\rq\rq{} structure  seems to persist in the overstretched state for which \lq\lq{}the most straightforward explanation ... is that the new structure, generated during overstretching at 110 pN, consists of two single DNA strands lacking hydrogen bonds between the bases, wrapped around each other with a linking number close to that of relaxed dsDNA\rq\rq{}\cite{vanMam2009}.

The quantity of interest in this context is the entropy difference   between the molten state with the geometrical constraint and the one without it, or,
in other words, the {\em entropic penalty}  $\Delta S^{*}<0$ required to maintain the molten DNA with the inherited plectonemic structure, compared to the reference state of the two free strands. Geometrically constrained force-induced  melting does not release the full amount of DNA melting entropy $\Delta S=12.5 \> k_B$ (cf. previous subsection), but a reduced amount $\Delta S + \Delta S^{*}$. This modifies the relative thermodynamic stability of the double helix (\ref{eq:dhstab}), which now reads
\begin{equation}
\label{eq:dhstabmod}
\Delta G = ( T-T_m)  \Delta S  +  T \Delta S^{*} \quad
\end{equation}
per base pair, and results  in an {\em enhanced stability} of the double helix. Overstretching will now occur  when the elastic free energy difference compensates (\ref{eq:dhstabmod}).

In what follows I  will calculate the entropic penalty associated with the plectonemic structure  in two different ways, based on the FJC and the KP model, respectively. 

\subsubsection{FJC estimate of the constraint entropic penalty} 

The overstretched DNA with the plectonemic structure can be visualized as consisting of the two strands joined at successive points with a  spacing $10 a$, equal to the pitch of the original double helix. In the FJC model this corresponds to $\nu=10 a^{\prime}/ d$ segments in each strand, where $a^{\prime}=0.63$ nm and $d =2 \lambda=$ 2.0  nm, are,  respectively,  the monomer distance and the Kuhn length of ss-DNA. The total number of such plectonemic joints will be $N/10$.

Neglecting any effects arising from excluded volume, the probability of two distinct, partially stretched strand pieces, each with $\nu$ elements,  starting from a common origin and meeting at a certain point  ${\vec r}$  in space within an axial distance $b$ and a radial distance $\rho$, is $p_{\nu}^2$, where
\begin{equation}
   p_{\nu} =  \pi  b \rho^2  \> P_{\nu}  ({\vec r}) 
   \label{eq:probplect}
\end{equation}
and 
\begin{equation}
 P_{\nu} ({\vec r}) = \left(\frac{3}{2\pi \nu d^2  }\right)^{3/2} e^{-\frac{3r^2}{2\nu d^2}} \left(1 -\frac{3}{4\nu} + \cdots \right)
\end{equation}
is the vector end-to-end distance probability distribution of the FJC chain.

The conformation with all $N/10$ plectonemic joints has a probability of occurrence $(p_{\nu}^2)^{N/10}$, i.e. an entropic penalty $\Delta S^{*} = k_B \ln p_{\nu} /5$ per original base pair. Inserting some plausible values for the contact in terms of the inherited double helix, e.g. $b=\rho=a/2$, results in  an estimate of $\Delta S^{*} = -2.16 \> k_B$.

\subsubsection{KP estimate of the constraint entropic penalty} 

The calculation is quite analogous with that of the FJC, the only difference now being that $\nu=10$ and
\begin{equation}
   p_{\nu} = \frac{ \pi  b \rho^2 }{ a^{\prime 3}} \> P_{\nu} \left (10 \frac{a}{a^{\prime}}\right) 
   \label{eq:probplectKP}
\end{equation}
where $P_{\nu}(x)$ is now the vector end-to-end distance distribution function, measured in units of the segment length, computed for the KP model with the stiffness parameter of ss-DNA (cf. section \ref{sec:DNAfe}).

Computation of the KP end-to-end distribution function can be performed, for relatively small systems, by direct matrix multiplication and Fourier transform  inversion \cite{Errami2007, Marko2005}. Noting that the Fourier transform of the end-to-end distribution function, $P_N(q)$, can be formally obtained from ${\hat Z}_N(f)$ (cf. (\ref{eq:ZKPf}) and (\ref {eq:Zfr}) above) by the substitution $\beta f \to iq$, it is possible to write
\[
P_N(q)=   ({\bf {\hat U}}^N)_{00}  \quad,
\]
where the matrix ${\hat U}$ is defined as in (\ref{eq:tll}), with $F$ substituted by 
\begin{eqnarray}
	{\hat F}_{ll'}({\tilde f})   	&	= & \int_{-1}^{1} \> dx P_l(x)P_{l^\prime}(x)e^{ i qa x}  \\ \nonumber
	 &= &\sum_{k=|l-l'|,k+l+l'=2r}^{l+l'}  (2k+1) (-i)^k
	    \frac{1}{r+1/2}\\  \nonumber
	  &   & \cdot  \frac{\Psi(r-k)\Psi(r-l)\Psi(r-l')}{\Psi(r)} 
	    j_k(qa)  \quad,
\label{eq:fllclosed_Pq}	    
\end{eqnarray}
where $j_k$ are now spherical Bessel functions, e.g. $j_0 = \sin x /x$.

The result of the numerical computation, for $r/a^{\prime}=10 a/a^{\prime}=5.4$, is $P_{10} = 5.50\times 10^{-4}$, leading, for the same values of the parameters $b$ and $\rho$ as in the FJC case, to an entropic penalty of $\Delta S^{*} = -2.06 \> k_B$ per original base pair. 

\subsubsection{Estimation of the overstretching threshold}
\begin{figure}[h!]
\includegraphics[width=0.4\textwidth]{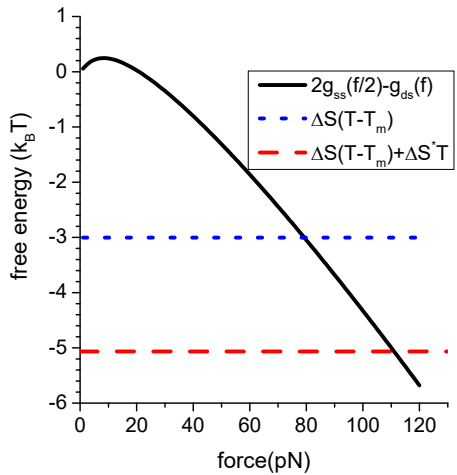}
\vskip -1.5truecm
\caption{Free energy balance resulting in DNA overstretching (constrained case).The solid line denotes the difference in elastic free energy between single- and double-stranded state. The dotted line denotes the duplex stability free energy at 20 C. The dashed line is the sum of duplex stability and entropic penalty 
$\Delta S=-2.06 \> k_B$ due to the constraint (KP calculation). Free energies are in units of $k_B T$. The intersection of dashed snd full line defines the overstretch threshold at 111  pN.
}
\label{fig:overstretch2}
\end{figure}

Fig. \ref{fig:overstretch2} displays the difference $2 g_{ss}(f/2) - g_{ds}(f)$ in the elastic free energies as a function of the applied force. Also shown is the stability free energy of the double helix, with and without the entropic penalty of the constraint, as given by  (\ref{eq:dhstabmod})   and (\ref{eq:dhstab}), respectively. Both effects lead to an enhanced stability of double-stranded DNA, in comparison with the unconstrained geometry. The overstretching force threshold, as estimated from the KP model calculation, with lies at 111 pN, in excellent agreement with experiment \cite{vanMam2009}.

\section{Discussion}

The present analysis of the overstretching transition involves a number of  material parameters for both ds- and ss-DNA. It is important to keep in mind that only two of these parameters originate in fitting the data of the experimental force-extension curve, namely the persistence lengths. All others are either standard accepted values (e.g. the crystallographically obtained values for the monomer distances $a$ and $a^{\prime}$, melting temperature $T_m$ and melting entropy $\Delta S$ per base pair) or, at the very least (e.g. elastic stretch moduli for ss-DNA and ds-DNA), physically plausible published values.  Not surprisingly, the limited space of the remaining free parameters improves the quality of the estimates obtained. Both persistence lengths are consistent with the values obtained by other researchers. 

In particular the value $\lambda =1$ nm obtained here for the persistence length of ss-DNA deserves a brief comment. The value lies at the low end of current estimates of 0.75 - 2 nm. Specifically, it compares well with the zero loop origami-based estimates of \cite{Roth2018}, in accordance  with the expectation of loops unwinding under the influence of the external force. On the other hand, it is significantly lower than the SAXS result of \cite{LipfertDoniach2012}, 1.7 nm at 150 mM NaCl concentration. Although a small part of the discrepancy may be accounted for by the use of the WLC continuum model in \cite{LipfertDoniach2012}, it is instructive to consider the physics underlying the different experiments. The partially - or fully - stretched state of \cite{vanMam2009},  is probably far less responsive to the electrostatic repulsion forces acting between segments. As a result, observed persistence lengths may be lower than those observed in force-free solutions and actually approach values corresponding to high salt concentrations, where  screening of electrostatic repulsion occurs. Independent sets of measurements \cite{LipfertDoniach2012, Chen2012} suggest that there is an intrinsic lower limit of ss-DNA persistence length, of approximately 1 nm and 0.75 nm, respectively, which is approached at high salt concentrations. It is quite possible that experiments performed near the stretched state effectively detect this limit of extreme chain flexibility.

The results of the thermodynamic analysis confirm the conjecture made \cite{vanMam2009} about the plectonemic nature of the overstretched state in the constrained geometry. It should be noted that this conclusion does not depend on the detailed choice of geometrical parameters specifying what exactly is considered as a contact. For example, increasing the linear dimension of a contact from $a/2$ to $a$ in both the radial and axial directions would result in a decreased entropy penalty by $ (1/5)\ln 8 =0.41 k_B$, changing the predicted threshold to $105 $ pN.

Are there any plausible alternatives to the plectonemic structure of the geometrically constrained molten state? A radically weaker variant would be to consider solely the entropic effect of holding the ends of both strands stretched, without imposing any further structural motifs. Such an alternative calculation results in a much smaller entropic penalty of the order $-0.18 k_B$ per base pair, roughly ten times less than that of the plectonemic structure, and cannot account for the observed force threshold. In other words, entropic considerations effectively demand some sort of internal structure in the molten state. The \lq\lq{}inherited\rq\rq{} structure with knots following the pitch of the double helix, as conjectured in \cite{vanMam2009} is probably the simplest model satisfying this demand.

I would like to thank Michel Peyrard for reading an earlier version of the manuscript. 

I acknowledge support by the project \lq\lq{}Advanced Materials and Devices (MIS 5002409)\rq\rq{}, implemented under the “Action for the Strategic Development on the Research and Technological Sector”, funded by the Operational Program \lq\lq{}Competitiveness, Entrepreneurship and Innovation (NSRF 2014-2020)\rq\rq{} and cofinanced by Greece and the European Union  (European Regional Development Fund).

\bibliographystyle{aps}
\bibliography{StatPhysDNAbibliography}







\end{document}